\documentclass{article}
\usepackage[utf8]{inputenc}
\usepackage{amsmath}
\usepackage{geometry}
\usepackage{wrapfig}
\usepackage{graphicx}
\usepackage{float}
\usepackage[superscript,biblabel]{cite}
\usepackage[font=small]{}
\usepackage{lineno}
\title{Structural hierarchy confers error tolerance in biological tissues}
\author{Jonathan Michel and Peter Yunker}

\begin{document}

\maketitle

\begin{center}
\section*{Abstract}
\end{center}
\textbf{
Structural hierarchy, in which materials possess distinct features on multiple length scales, is ubiquitous in nature; diverse biological materials, such as bone, cellulose, and muscle, have as many as ten hierarchical levels \cite{fratzl, meyers, launey, yao}. Structural hierarchy confers many mechanical advantages, including improved toughness and economy of material \cite{lakes, wegst}. However, it also presents a problem: each hierarchical level adds a new source of assembly errors, and substantially increases the information required for proper assembly. This seems to conflict with the prevalence of naturally occurring hierarchical structures, suggesting that a common mechanical source of hierarchical robustness may exist. However, our ability to identify such a unifying phenomenon is limited by the lack of a general mechanical framework for structures exhibiting organization on disparate length scales. Here, we use simulations to substantiate a generalized model for the tensile stiffness of hierarchical, stretching-stabilized, filamentous networks with a nested, dilute triangular lattice structure. Following seminal work by Maxwell and others on criteria for stiff frames\cite{jmax,feng,das,mao} , we extend the concept of connectivity in network mechanics, and find a mathematically similar dependence of material stiffness upon each hierarchical level. Using this model, we find that the stiffness of such networks becomes \textit{less} sensitive to errors in assembly with additional levels of hierarchy; though surprising, we show that this result is analytically predictable from first principles, and thus likely model-independent. More broadly, this work helps account for the success of hierarchical, filamentous materials in biology and materials design, and offers a heuristic for ensuring that desired material properties are achieved within the required tolerance.
}

\begin{center}
\item \section{Introduction}
\end{center}
Living systems organize across many distinct levels, spanning from molecular to macroscopic scales. Such hierarchical arrangements endow organisms with many beneficial material properties; they may have high strength-to-weight ratios, exhibit strain stiffening, or be robust against fracture \cite{fratzl, meyers, launey, yao, sen}. A seeming drawback of this approach, however, is the enormous amount of information needed to specify the structure of highly hierarchical tissues and the increased number of opportunities for stochastic errors. Even for a self-assembled material, each hierarchical level increases the number of local minima in the free energy landscape, increasing the opportunity for kinetic errors in assembly \cite{hormoz, zeravcic}. While one may reasonably fear that this cascade of errors will undermine the reliable realization of self-assembled hierarchical materials, structural hierarchy is employed effectively by organisms belonging to many diverse evolutionary lineages \cite{meyers, dunlop, harris}. Such widespread success suggests the presence of an underlying mechanism responsible for this emergent robustness. However, the number of elements necessary to describe a hierarchical structure grows geometrically with the number of hierarchical levels; thus, a ten-level structure is computationally inaccessible. While identification of the underlying principles responsible for hierarchical robustness would greatly aid in explaining the ubiquity of natural hierarchical structures, this objective first requires developing a mechanistic understanding of how each scale contributes to a material's overall properties.
\par
To gain a foothold in the study of hierarchical materials mechanics, we focus on a highly tractable model system: a triangular lattice of nodes connected by harmonic springs. Frames made of slender, elastic beams have long been of interest in technical mechanics \cite{lakes,jmax,feng,das,mao,gab,wilhelm,fengj} and the physics of living tissue \cite{fratzl, meyers, launey, dunlop, head, lind, broed2,amuasi,licup}, and recent work has demonstrated that fibers can generically emerge from diverse building blocks \cite{lenz}. Further, the mechanics of elastic networks are easily interpretable through the Maxwell counting heuristic; briefly, to constrain every degree of freedom in the network, there must be $2d$ bonds per node, where $d$ is the dimensionality of the system \cite{jmax,broed1}. While much work has been done to characterize elastic networks constructed with a single important length scale \cite{feng, gab, das, mao, zhang, lub}, we lack a general characterization of elastic networks constructed with multiple disparate length scales; \textit{a priori}, it is unclear how Maxwell counting applies to hierarchical structures. Are there distinct degrees of freedom associated with `large' nodes, just as there are with `small' nodes? How do constraints on large and small length scales compare? Identification of underlying mechanisms that make hierarchical structures robust first requires developing a comprehensible hierarchical model.
\par

Here, we introduce a model system with a nested, dilute triangular lattice structure, in which distinct network connectivities can be defined on multiple scales. We examine the dependence of tensile stiffness on each of these connectivities, and capture this relationship with a simple model. Using this model, we then assess the resilience of a hierarchical material's mechanical properties in the presence of random errors in assembly.

\begin{center}
\item \section{Description of Model System}
\end{center}
\subsection{Geometrical Characteristics}

We consider an extension of the well-studied dilute, triangular lattice in two dimensions. Nodes arranged in a triangular Bravais lattice are connected to nearest neighbors, and bonds are then removed at random such that some fraction, referred to as the bond portion, $p$, remains. The infinite triangular lattice has a connectivity of 6 bonds per node when $p = 1$, while in two dimensions Maxwell counting dictates a minimum connectivity of $4$ bonds per node; thus, the infinite, dilute triangular lattice should lose stiffness when $p$ falls below $\frac{2}{3}$. Lattices of finite size would require a slightly higher bond portion, due to the presence of under-constrained nodes at the boundary. This prediction has been thoroughly confirmed for ball-and-spring networks, via simulation and mean field-theoretic approaches \cite{feng, gab, das, mao}.

\par
We create hierarchical triangular lattices through an iterative process, in which the bonds of the lattice at one length scale are in turn crafted from smaller-scale triangular lattices. In principle, this process can be iterated \textit{ad infinitum}; in practice, if the number of large bonds is held constant, the total number of nodes grows geometrically with the number of hierarchical levels. This places a practical limit on the number of levels that can be considered in simulations. We have numerically constructed and simulated lattices with 1, 2, and 3 levels of structural hierarchy (figure 1). Bond portion can be independently set on each level of a hierarchical network.

\begin{figure}
\centering
\includegraphics[width=.9\textwidth]{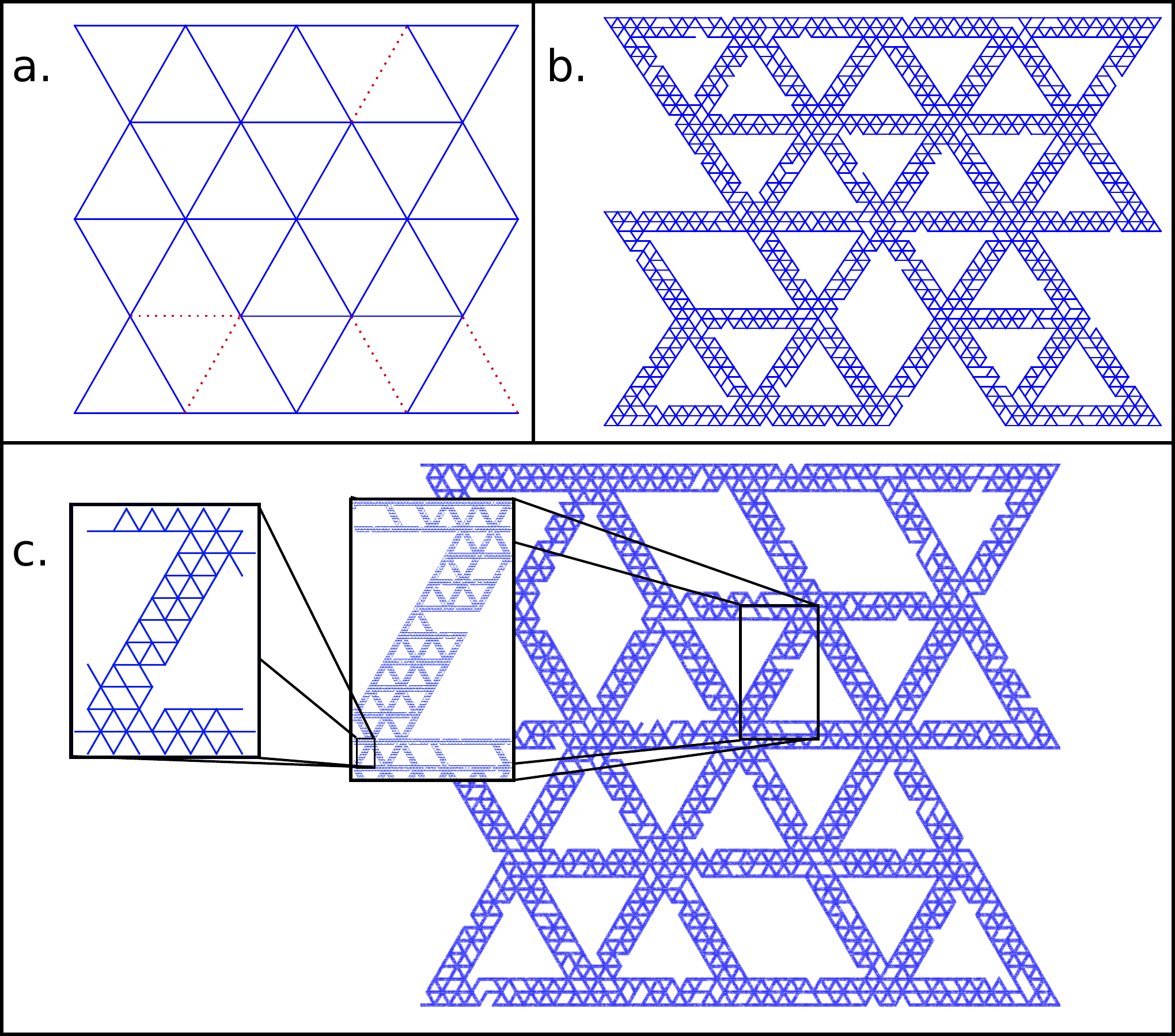}
\caption{\textbf{a.} A dilute triangular lattice with one levels of structure. Missing bonds are indicated with dashed lines. \textbf{b.} The same dilute triangular lattice, with each bond replaced by a smaller-scale, dilute triangular lattice \textbf{c.} An extension to three levels of structural hierarchy}
\end{figure}

\section{Investigation of Stiffness}
\subsection{Hierarchical Model of Stiffness}
We propose to model the stiffness of our networks by generalizing the scaling law proposed by Gaborczi, \textit{et al.}, for a single-scale, diluted triangular lattice \cite{gab}. Gaborczi, \textit{et al.}, found that, for the ball-and-spring case, components of the elastic constant tensor should have the form

\begin{equation}
K = 
\begin{cases}
\label{sls}
k\frac{p - p_c}{1-p_c} \text{,} p \geq p_c\\
0 , p < p_c
\end{cases}
\end{equation}

where $p_c$ is the minimum bond portion necessary for marginal stiffness, and k is the value of the modulus when the network is fully connected. For the infinite triangular lattice, $p_c = \frac{2}{3}$.
We propose to describe large-scale bonds using an effective stiffness with the form of ~\eqref{sls}, and introduce $p_{large}$ and $p_{small}$, the portion of bonds retained on the large and small scales. Because of the finite width of large-scale bonds, we will not assume the minimum small- or large-scale bond portions needed for marginal stiffness are $\frac{2}{3}$. We then conjecture that the stiffness of a large scale bond is inherited from its small scale structure, such that the overall stiffness scales as:

\begin{equation}
K = k\frac{\left(p_{large} - p_{c, large}\right)\left(p_{small} - p_{c, small}\right)}{\left(1-p_{c,large}\right)\left(1-p_{c,small}\right)}
\end{equation}

where $K$ is tensile stiffness and $k$ is the stiffness for a network fully connected on all scales. We anticipate that, for arbitrary levels of structural hierarchy, stiffness will follow the general form

\begin{equation}
\label{general}
K = k\prod_{i = 1}^N \frac{p_i - p_{c, i}}{1-p_{c,i}}
\end{equation}

for some general number $N$ levels of structural hierarchy.

\subsection{Simulation Procedure}
Networks were simulated in two dimensions, with ball-and-spring interactions between pairs of connected nodes. Nodes along the tops of networks were uniformly displaced along the vertical direction, after which the y coordinates of the top and bottom nodes were fixed. Next, the x coordinates of top and bottom nodes, as well as both coordinates of all other points, were relaxed using the FIRE algorithm \cite{fire}. A uniform stretching modulus of unity was assigned to each bond, and nodes were relaxed until the RMS residual force in the network was less than $1 \times 10^{-10}$ in units of stretching modulus in the case of the one and two-level networks, and $1 \times 10^{-9}$ in units of stretching modulus for three level networks. We extract the tensile stiffness by fitting plots of elastic energy versus strain to parabolas.

\begin{figure}[H]
\centering
\includegraphics[width=.6\paperwidth]{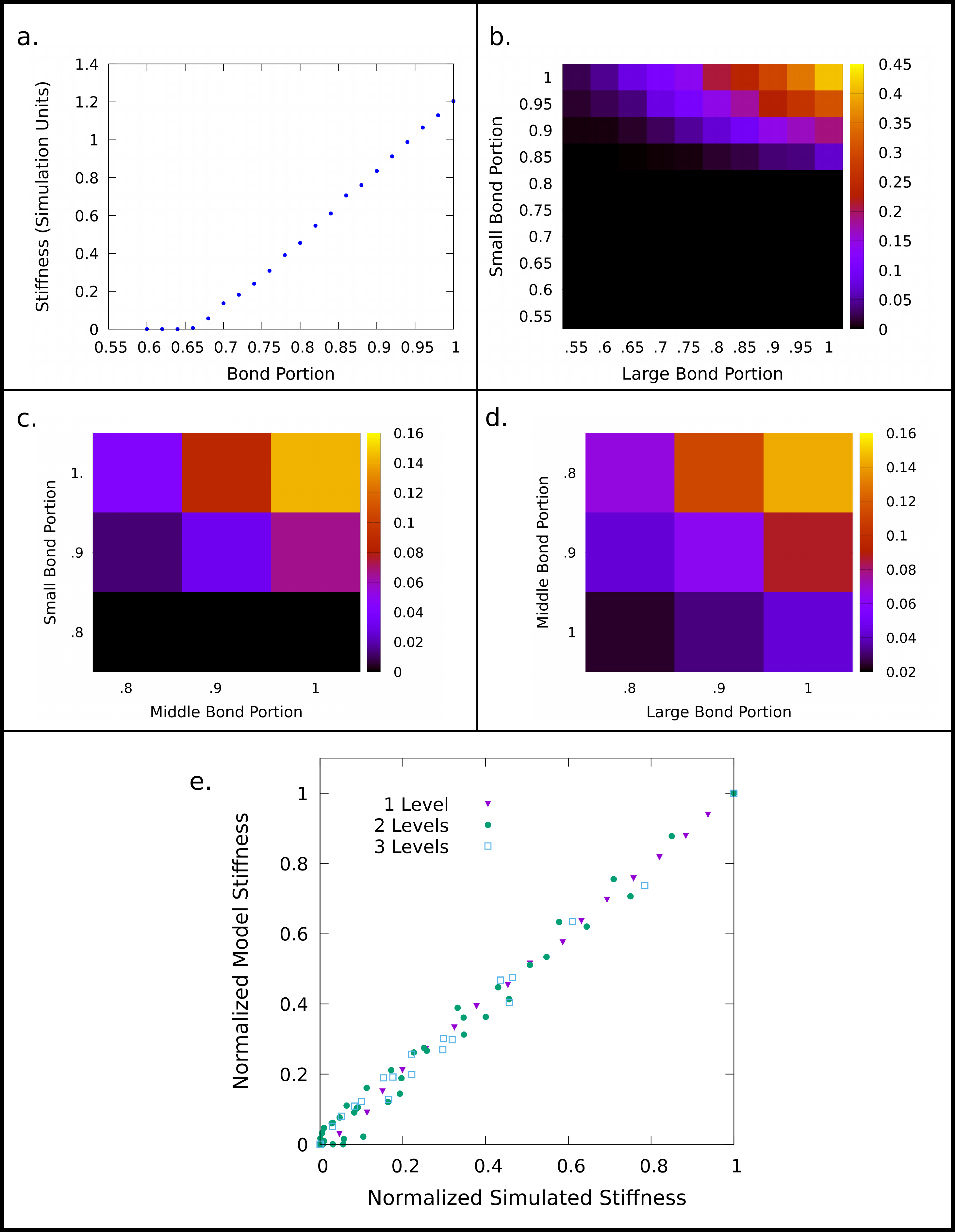}
\caption{\textbf{a}. Simulated stiffness plotted vs. bond portion for one length scale. \textbf{b}. Simulated stiffness vs. small and large bond portion for a two-level network. \textbf{c} A slice in bond portion space for three level networks with full connectivity on the largest scale. \textbf{d} A slice in bond portion space for three level networks with full connectivity on the smallest scale. \textbf{e.} Stiffness normalized by maximum stiffness for networks with one, two and three levels of structure.}
\end{figure}

\subsection{Simulation Results}
We simulated 1-, 2-, and 3-level lattices with a wide range of bond portions, and measured their stiffness (Fig. 2). Before comparing to our conjectured model, we must identify each critical bond portion for 1-, 2-, and 3-level lattices. For the 1-level lattice, we recover the expected linear relationship between stiffness and bond portion (Fig. 2a). We find the critical bond portion to be 0.670.

For the 2-level lattice, we find that stiffness increases as either the small or large bond portion is increased (Fig. 2b). We extract the critical bond portion on the small (large) scale by setting the large (small) scale bond portion to 1.0, and identifying the small (large) scale bond portion at which the stiffness is zero (see Supplemental Information). We find the critical bond portions are 0.83 and 0.60 for the small and large scales, respectively. Initially, it may be surprising to find that the critical bond portion on the large scale is less than 0.67. However, as large scale bonds are endowed with a finer scale structure, they acquire an effective bending stiffness, rather than being governed strictly by harmonic, central force interactions. Notably, networks with bonds possessing bending stiffness have been demonstrated to be rigid even below the classic isostatic point ~\cite{das,mao}. Thus, it is crucial that $p_c$ be directly measured, and not assumed from Maxwell counting.

For the 3-level lattice, we find that stiffness increases as any bond portion is increased (Fig. 2c and d and Supplemental Information). Following the same approach used for the 2-level lattice, we find the critical bond portions are 0.83, 0.73, and 0.62 for the small, medium, and large scales, respectively. 

To test our conjectured model, we compare the stiffness measured in simulations to the stiffness predicted by our model. There are no free parameters in our model; we normalize stiffnesses by the maximum values for 1-, 2-, or 3-level lattices, and use the identified critical bond portions. We find remarkable agreement; linear fits between simulated and predicted quantities have $r^2$ values and slopes, respectively, of $0.983$ and $1.003$ for one level, $0.989$ and $0.995$ for two levels, $0.978$ and $0.992$ for three levels, and $.998$ and $.994$ for all data combined (Fig. 2e).

Thus, our conjectured formulation for stiffness of a hierarchical lattice accurately describes 1-, 2-, and 3-level lattices. Crucially, this model suggests that the smallest length scale can always be replaced by a network of even smaller bonds, and the stiffness will remain the product of all excess bond portions. This general formulation now facilitates investigation of highly hierarchical (e.g., 10-level) lattices.

\section{Error Tolerance}

Now that we have obtained a general relationship for the stiffness of a hierarchical structure, we are primed to consider the possibility of random errors in assembly, a likely complication in any real assembly process. We focus in particular on how stochastic deviation from some target set of connectivities (on all relevant length scales) results in a deviation in network stiffness. We consider two distinct regimes. In the first case, we consider a target point near the isosurface along which stiffness vanishes. In the second, we consider a target point in bond portion space far from both the limiting case of full connectivity on any length scale and from the contour of vanishing stiffness.

We first consider target points in bond portion space corresponding to marginally stiff structures. Such points are of interest, as highly compliant materials have critical biological roles \cite{fratzl}, and are attracting increasing attention for technological applications \cite{wegst}. Such materials typically must not be susceptible to critical transitions in their elastic moduli as a result of small fluctuations in their fine-scale structure. 

We utilize a numerical technique to estimate the distribution of stiffness arising from random errors. First we choose a nominal point in bond portion space, then we add Gaussian random noise to the bond portion on each length scale. The stiffness of the resulting ``noisy'' point is then estimated by means of an interpolated function computed from simulation data for 1- and 2-level lattices, and a fitted model for 3-level lattices (Fig. 2); for lattices with more than 3 levels, we use equation~\eqref{general}. This process is carried out for 50,000 trials.

Strikingly, the variance in stiffness is greatly reduced with each additional level of hierarchical structure (Fig. 3a). Further, stiffness distributions for the single-level and two-level networks exhibit a large peak at zero, which is absent in the stiffness distribution of the three-level network. Thus, despite having a much higher error rate, the three-level network is more reliably constructed than the one-level network, and can more readily avoid stochastically generating a floppy network.

\begin{figure}
\centering
\includegraphics[width=0.5\paperwidth]{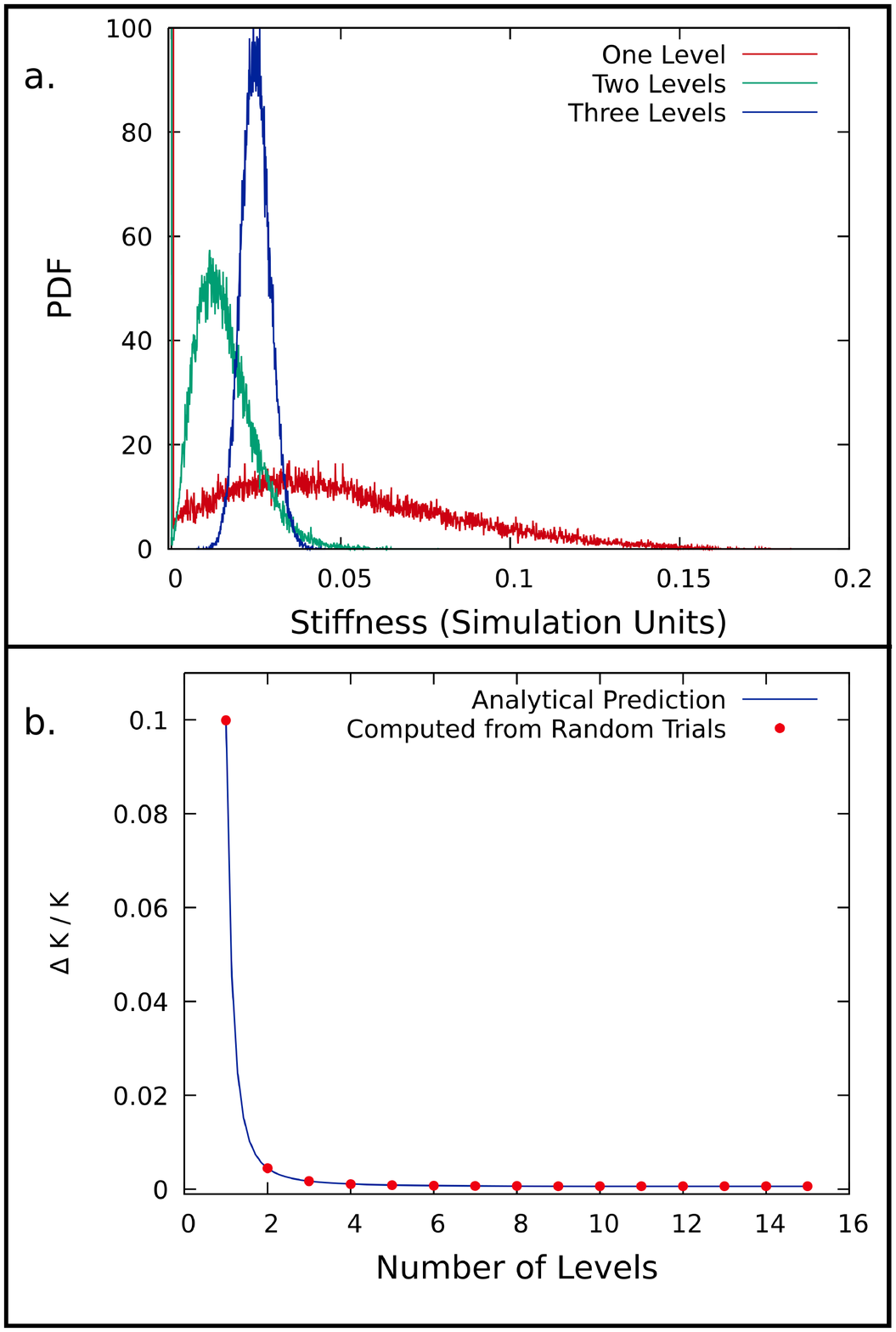}
\caption{a. Stiffness probability distribution functions estimated from histogram data for networks with one, two, and three levels of structural hierarchy. Points in one, two, and three-dimensional bond portion with the same minute nominal stiffness were chosen, and Gaussian random variables were added to each bond portion. Note the spike in the PDFs for one and two-level networks at zero stiffness.  b. Relative error in stiffness vs. levels of hierarchy is plotted for $\bar{K} = .001$, $\sigma = .0001$. As additional levels of structural hierarchy are added, the relative error in the tensile stiffness decreases precipitously at first, and the effect saturates at a certain number of levels. Provided the assumptions leading to equation ~\eqref{scale_law} hold, our analytical theory and numerical approach are in close agreement ($r^2 \approx 1$). Here, the product of excess bond portions is $.001$, and the noise has amplitude $.0001$ on each scale.}
\end{figure}

Next, we seek a general understanding of target points far from any boundary in bond portion space (see supplement for detailed derivation). For $N$ levels, there is a nominal excess bond portion, $p_e$, for each level; we consider identically and independently distributed deviations from the nominal bond portions according to a normal distribution with zero mean and standard deviation $\sigma$. Referring to equation ~\eqref{general}, we define the reduced stiffness:
\begin{equation}
\bar{K} = \frac{K}{k \prod_{i=1}^N \left(1 - p_{c,i}\right)}
\end{equation}
With this definition we find the expected deviation in the stiffness of a network with $N$ hierarchical levels to be: 

\begin{equation}
\label{scale_law}
\frac{\Delta \bar{K}}{\bar{K}} \approx \frac{\sqrt{N}\sigma}{\bar{K}^{1/N}}
\end{equation}

where the approximation holds when $\sigma \ll p_e$. For a target $\bar{K}$, this
functional form predicts the optimal number of levels to be

\begin{equation}
N^* = \lfloor -2 \ln\left(\bar{K}\right)\rfloor
\end{equation}

To test our analytical result, we again utilize the above numerical approach to calculate standard deviation in stiffness for networks with one- to fifteen-hierarchical levels. For lattices with more than 3 levels, we use equation~\eqref{general} to estimate the stiffness of the resulting ``noisy'' point. We find very good agreement between the numerically generated data and our analytical prediction ($r^2 \approx 1$ for the case shown in figure 3b.).

\par
We briefly note that the results presented above do not strictly require identically and
independently distributed random errors in bond portion. Similar robustness against fluctuation in elastic moduli can occur for varying error rates on different scales,
and for distributions of random errors in which errors in bond portion on different length
scales are correlated (see Supplementary Information for an in depth treatment). Interestingly, investigating networks with different error rate distributions allows us to identify a useful heuristic. Consider a two-level network with the same error rate in its large-scale bond portion as a one-level network. The two-level network will have a smaller variance in its stifness than the one-level network as long as its small bond portion error rate is less than three times larger than its large bond portion error rate.
\section{Discussion}
Contrary to expectation, the elastic moduli of hierarchical materials are more reliably controlled than the elastic moduli of materials with one relevant length scale. Thus, it is actually easier to make hierarchical structures than to make single-length scale structures. This finding may have wide-ranging implications for evolutionary biology and materials design. 
\par
It may seem that evolving progressively larger, more complex bodies is accompanied, and impeded, by a growing assortment of mechanical challenges. To the contrary, this work provides evidence that adding hierarchical complexity can actually \textit{reduce} stochastic variation in material properties. This effect decreases the need for co-evolving error correcting mechanisms, thus facilitating the evolution of new traits that are `good enough' for an organism to survive.
\par
Upon successful assembly, structural hierarchy is known to endow materials with a host of desirable properties that are unattainable with single-length scale structures. Our finding that hierarchy also reduces susceptibility to stochastic errors suggests bottom-up production processes for synthetic hierarchical materials, in which modest, but nonetheless discernible errors occur at each stage, may nonetheless produce a finished product which performs as intended.
\par
More generally, we have broadened the scope of Maxwell's visionary means of characterizing frames, to allow for quantitative understanding in cases in which it has proven difficult to relate materials' emergent properties to their fine-scale structure. While recent computational advancements have enabled study of biological macromolecules over experimentally relevant time scales \cite{JC}, comprehensive understanding at the level of an organism demands a coarse-graining procedure for which our model may offer a useful road map. A generalized counting heuristic may also offer a means of expediting feasible, yet cumbersome calculations in materials design. Hierarchical materials necessarily have many design attributes, but our accessible model may considerably narrow the search of parameter space needed to reach a goal.

\section{Methods}

\subsection{Network Creation}

First, a large-scale lattice is created and diluted. Dilution begins with the shuffling of all bonds, after which a random minimum spanning tree is created using Kruskal's algorithm. Bonds are then drawn at random from those bonds not used to create the spanning tree until the desired bond portion is reached. Next, a small-scale lattice is overlaid such that the large scale bond length is an integer multiple of the small-scale bond length, and the position of each large-scale node coincides with the position of a small-scale node. Each small-scale bond lying within a large-scale bond is retained, after which small-scale bonds are diluted to the desired bond portion. Small-scale bond dilution is carried out in such a way that a system-spanning contact network remains, no large-scale bond is severed, and all adjacent large-scale bonds remain connected. This process may then be repeated, with the small-scale network taking the role of the large-scale  network. This process is described schematically in Figure 1.

\subsection{Finding the critical bond bond portions}
Critical bond portions for a network with $n$ levels are computed from simulation data for
each level by choosing all points in bond portion space for which the network is fully
connected on the other $n-1$ levels, and the stiffness is non-zero. Extrapolation is then used
to find the x intercept of a line fit to these points to determine the critical connectivity
for the isolated level.

\begin{figure}[H]
\includegraphics[width=.7\textwidth]{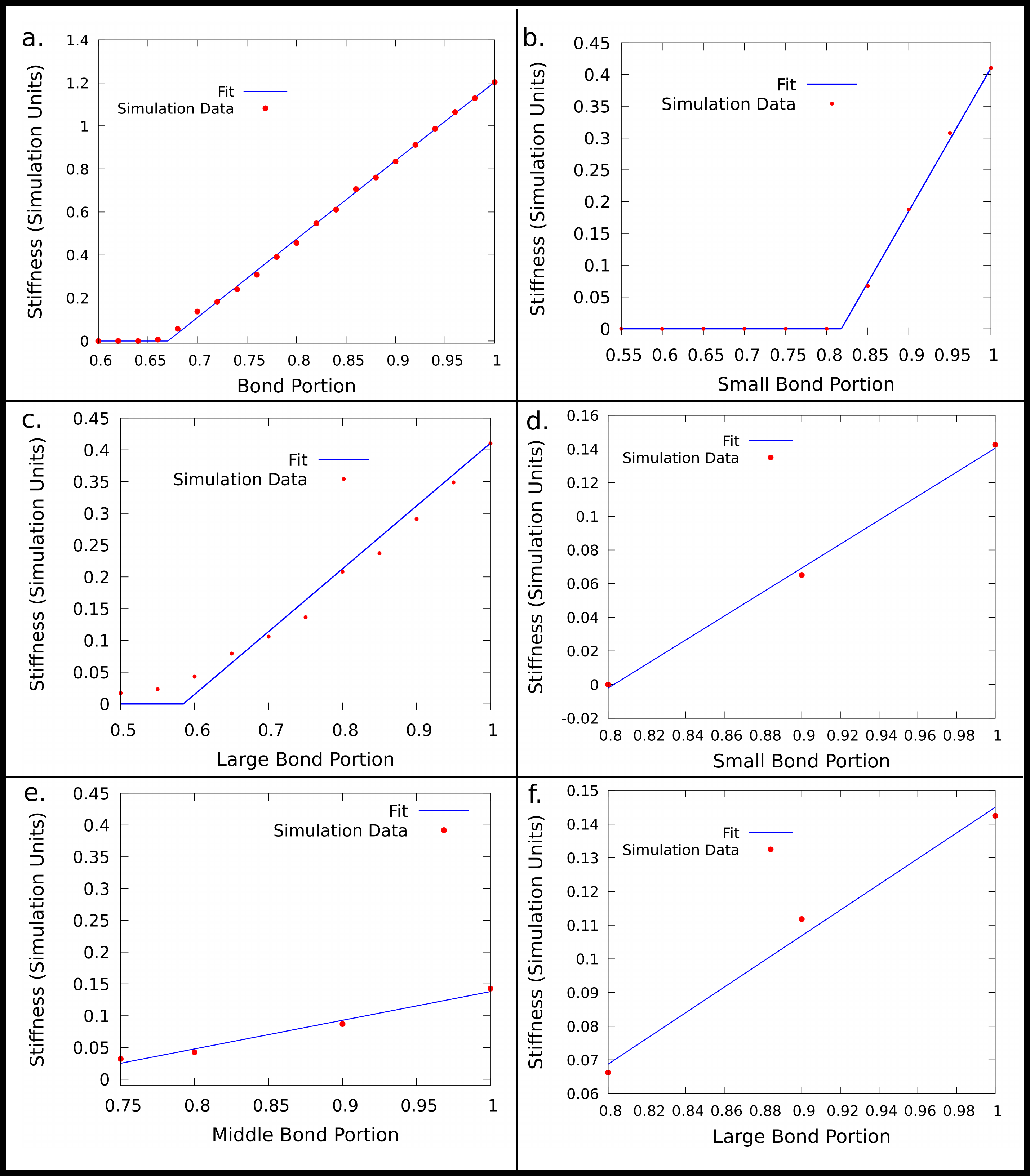}
\caption{Panel \textbf{a} illustrates the determination of the critical bond portion for one level. Panels \textbf{b} and \textbf{c} demonstrate the determination of the critical bond portion for the large and small-scale critical connectivities for the two-level case, and
panels \textbf{d}-\textbf{e} illustrate determination of critical connectivity for the
large, medium and small scales for the three-level case.}
\end{figure}

\subsection{Sampling of Bond Portion Space}
For one-level and two-level networks, we managed to sample bond portion space from $.55$ to $1$ in
increments of $.05$ for all levels. Owing to the computational expense of simulating
three-level networks, we sampled bond portion space from .8 to 1 in increments of .1 for all
levels, and additionally sampled points just above and just below the critical bond portion for 
each level. This limited interpolation of three-level networks to a cube in bond portion space
spanning the range from $.8$ to $1$ in each direction. As such, estimates of the distribution of
stiffness given a nominal point and an error rate were carried out using a fitted model with the
form of equation ~\eqref{general}.

\section{Supplementary Information}

\subsection{Analytical prediction of hierarchical robustness}

First, we consider a nominal point far from any boundary in bond portion space. For $N$ levels, we define an excess bond portion $p_{e,i}$ for each level, with $1 \leq i \leq N$:

\begin{equation*}
p_{e,i} = p_i - p_{c,i}
\end{equation*}

Now let the deviation from the nominal excess bond on the $i$th level be $\delta_i$, with
each $\delta_i$ identically and independently distributed according to a distribution $\mathcal{P}_i$ with zero mean and standard deviation $\sigma_i$, such that the total probability distribution function for a set of displacements $\left\{\delta_1, \dots, \delta_N\right\}$ is

\begin{equation*}
\mathcal{P} = \prod_{i=1}^N \mathcal{P}_i\left(\delta_i\right)
\end{equation*}

Referring to equation ~\eqref{general}, we define the reduced stiffness 
\begin{equation*}
\bar{K} = \frac{K}{k \prod_{i=1}^N \left(1 - p_{c,i}\right)}
\end{equation*}
With this definition we now find the expected deviation in the stiffness of a network with $N$ hierarchical levels. The mean of $\bar{K}$ is

\begin{align*}
\left<\bar{K}\right> &= \int \cdots \int \left( \prod_{i = 1}^N p_{e,i} + \delta_i\right) \mathcal{P}\left(\delta_1, \dots, \delta_N\right) d \delta_1 \cdots d \delta_N\\
&= \prod_{i = 1}^N p_{e,i}
\end{align*}

while the mean square stiffness is

\begin{align*}
\left<\bar{K}^2\right> &= \int \cdots \int \left(\prod_{i=1}^N \Big(\delta_i + p_{e,i}\Big)^2\right) \mathcal{P}\left(\delta_1, \cdots, \delta_N\right) d \delta_i \cdots d \delta_N\\
&= \prod_{i=1}^N \sigma_i^2 + p_{e,i}^2
\end{align*}

The standard deviation in stiffness is then

\begin{equation}
\label{sdev}
\Delta \bar{K} = \sqrt{\prod_{i=1}^N \sigma_i^2 + p_{e,i}^2 - \prod_{i = 1}^N p_{e,i}^2}
\end{equation}

Here it has been assumed that $p_{e,i} \gg \sigma_i$ for all $i$, so integration can be carried out with the assumption that equation ~\eqref{general} holds for all values $\delta_i$ that contribute appreciably to the integral. In the special case in which the excess bond portion is the same on each length scale, and
each bond portion has the same standard deviation, ~\eqref{sdev} reduces to
\begin{equation*}
\sqrt{\left(\sigma^2 + p_{e}^2\right)^N - p_{e}^{2N}}
\end{equation*}
, with $p_{e} = \bar{K}^{1/N}$. The relative error in stiffness then scales with $N$ as

\begin{align*}
\frac{\Delta \bar{K}}{\bar{K}} &= \frac{\sqrt{\left(\sigma^2 + p_e^2\right)^N - p_{e,i}^{2N}}}{p_e^N}\\
&= \sqrt{\left(1 + \frac{\sigma^2}{p_e^2}\right)^N - 1}\\
&\approx \frac{\sqrt{N}\sigma}{p_e}
\end{align*}

or
\begin{equation}
\label{uniform_dev}
\frac{\Delta \bar{K}}{\bar{K}} \approx \frac{\sqrt{N}\sigma}{\bar{K}^{1/N}}
\end{equation}

where the last approximation holds when $\sigma \ll p_e$. For a target $\bar{K}$, this
functional form predicts the optimal number of levels to be

\begin{equation}
N^* = \lfloor -2 \ln\left(\bar{K}\right)\rfloor
\end{equation}

\subsection{Accounting for Other Types of Error Distributions}
As mentioned in the main text, we also accounted for two additional classes of distributions of random errors in
assembly. In the first case, we still presume the errors to be independent on each scale and normally distributed, but
we allow the standard deviation of error in bond portion at the second length scale to be different from the standard deviation for all
other scales. For an $N$-level network, suppose $N-1$ levels exhibit random errors with standard deviation $\sigma_1$, while
the remaining level exhibits random errors in bond portion with standard deviation $\sigma_2$. In this case, the probability
distribution function for the $N$-dimensional vector of errors, $\vec{\delta}$, should take the form

\begin{equation}
\mathcal{P}\left(\vec{\delta}\right)  = \frac{1}{\left(2\pi\right)^{2/N}\sigma_1^{N-1}\sigma_2} \exp\left\{-\frac{1}{2 \sigma_1^2}\prod_{i\neq 2}\delta_i^2 - \frac{\delta_2^2}{2 \sigma_2^2}\right\}
\end{equation}

We consider once more a point in bond portion space with the same excess bond portion
on each level, and that this excess bond portion is much greater than either $\sigma_1$
or $\sigma_2$. A straightforward modification to the above derivation for a constant
standard deviation yields

\begin{equation}
\label{diff_dev}
\frac{\bar{K}}{\Delta \bar{K}} = \sqrt{\frac{\sigma_2^2}{\bar{K}^{2/N}} + (N-1)\frac{\sigma_1^2}{\bar{K}^{2/N}} + (N-1)\frac{\sigma_1^2\sigma_2^2}{\bar{K}^{4/N}}}
\end{equation}

We consider the greatest value for $\sigma_2$ for an $N$-level network such that the
relative variation in stiffness is no greater than the relative deviation in stiffness
for a single-level network with a bond portion distribution of width $\sigma_1$.
Equating the right-hand sides of ~\eqref{uniform_dev} and ~\eqref{diff_dev} yields

\begin{equation}
\frac{\sigma_2}{\sigma_1} = \bar{K}^{2/N}\sqrt{\frac{1}{\bar{K}^2} - \frac{N-1}{\bar{K}^{2/n}}}\sqrt{\frac{1}{\bar{K}^{2/N} + (N-1)\sigma_1^2}}
\end{equation}
We show this behavior below for the case in which the product of excess bond portions
is fixed at $.1$, and at all levels of structure but the second, the standard deviation
of the error in bond portion is $.001$. The ratio of the maximum standard deviation on the
second level such that the overall relative variation in stiffness remains less than or
equal to that for a one-level network is plotted vs. the number of levels of hierarchy.

\begin{figure}[H]
\includegraphics[width=\textwidth]{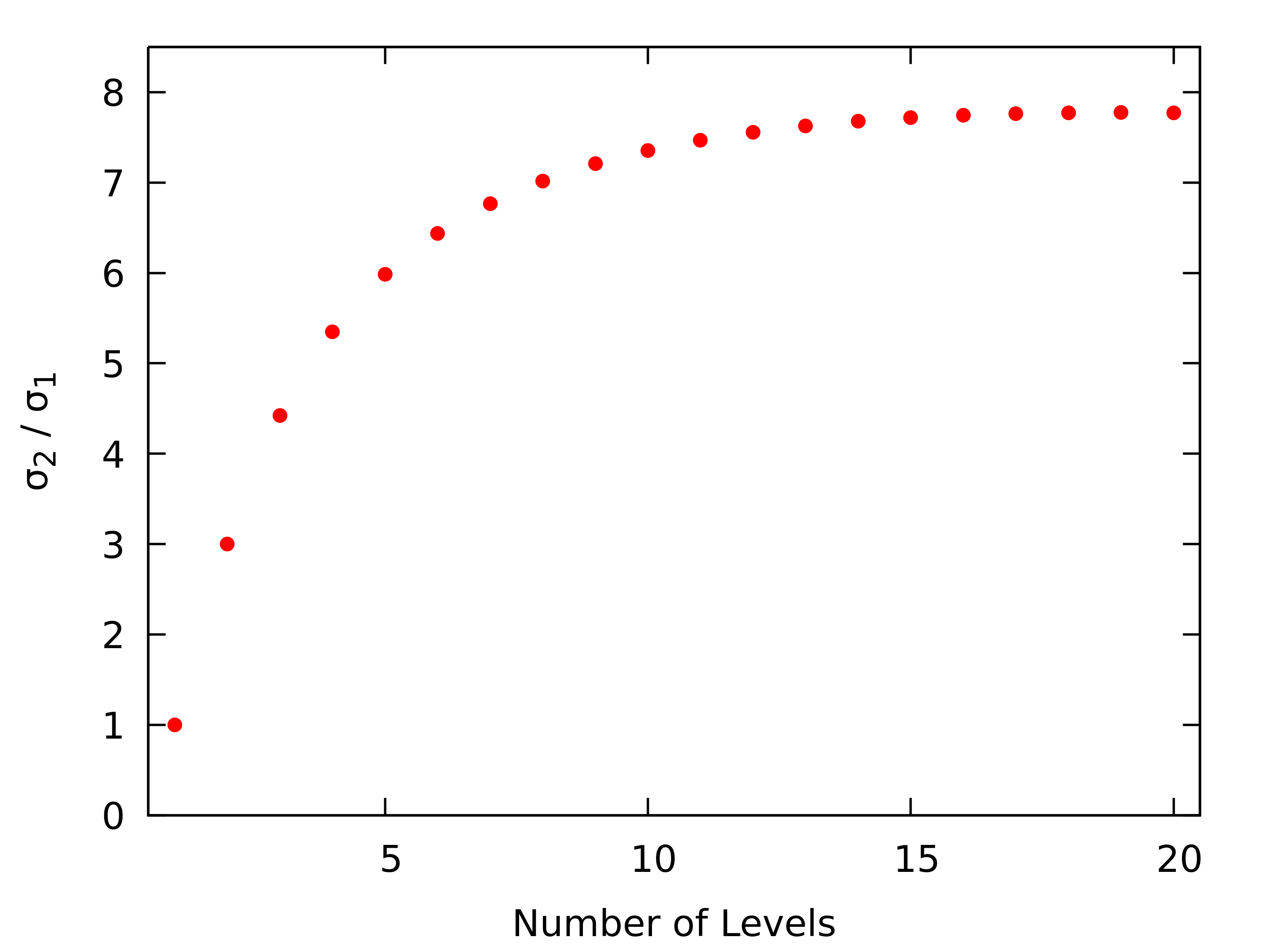}
\end{figure}

\par
We also consider the case in which the errors in bond portion on different
structural levels are identically distributed, but correlated. For a network with $N$
structural levels, we consider an $N \times N$ covariance matrix $\mathbf{\Sigma}$ which
takes the form

\begin{equation}
\Sigma_{i,j} = 
\begin{cases}
\sigma^2, & i=j\\
\rho \sigma^2, & i \neq j
\end{cases}
\end{equation}

It can be shown that
\begin{equation*}
|\mathbf{\Sigma}| = \sigma^N \left[1 + \left(N - 1\right)\rho\right]\left(1 - \rho\right)^{N-1}
\end{equation*}

In this case, the probability distribution function for a vector of bond portion errors
$\vec{\delta}$ is given by

\begin{equation*}
\mathcal{P}\left(\vec{\delta}\right) = \frac{1}{\left(2\pi\right)^{N/2}|\mathbf{\Sigma}|} \exp\left[-\frac{1}{2} \vec{\delta}^T \hspace{1pt} \mathbf{\Sigma^{-1}} \hspace{1pt} \vec{\delta}\right]
\end{equation*}

Below, we show a plot of relative error in stiffness vs. number of levels, with fixed $\sigma$ and varying coupling
strength $\rho$.

\begin{figure}[H]
\includegraphics[width=\textwidth]{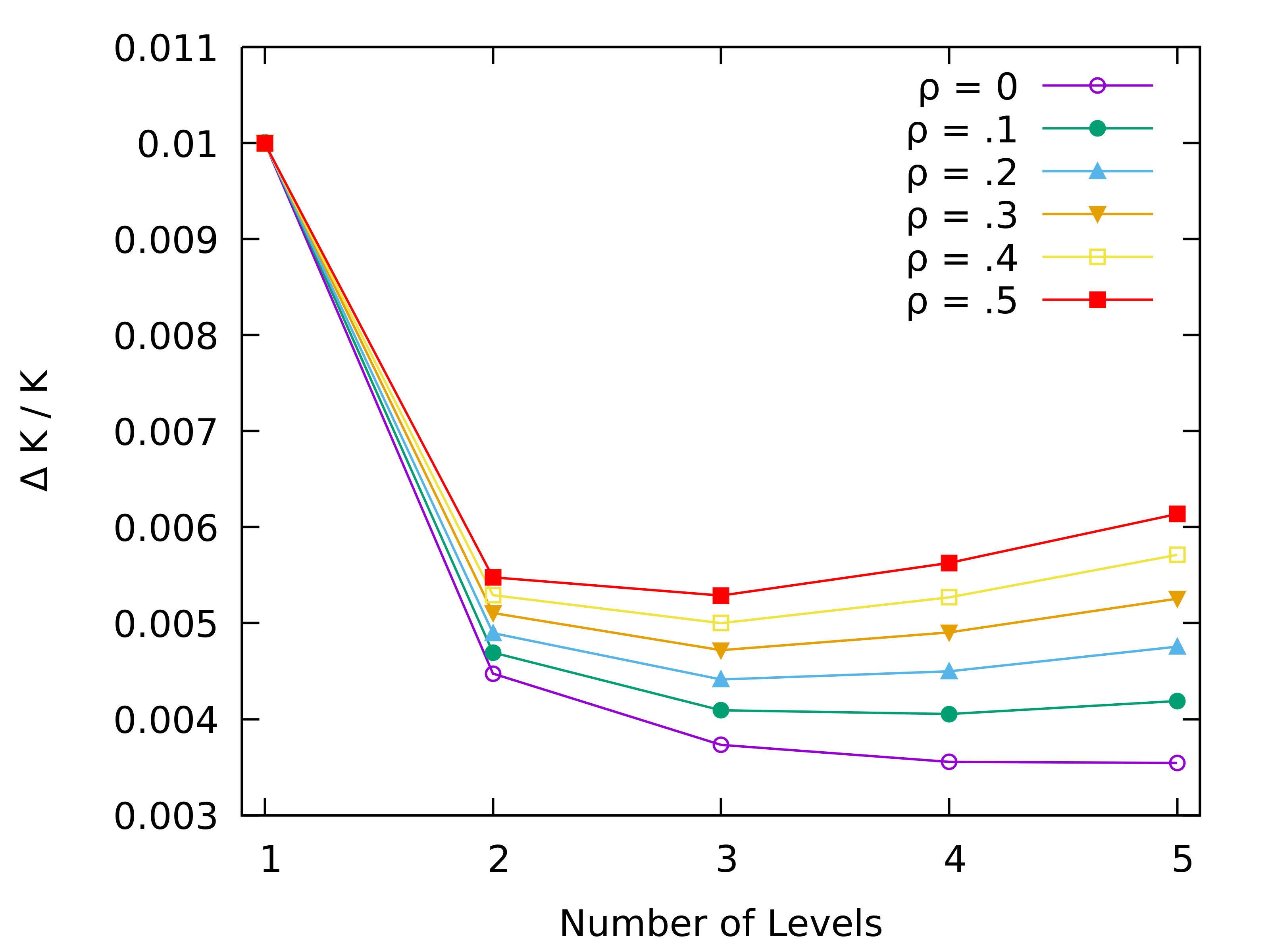}
\caption{The relative error is plotted against the number of hierarchical levels for
cases in which the product of excess bond portions is .1, and the diagonal elements
of the covariance matrix are $10^{-6}$, and $\rho$ is varied from 0 to .6.}
\end{figure}

In view of the results of these alternative investigations, we anticipate that protection
against fluctuation in stiffness is a generic benefit of structural hierarchy, and does
not depend sensitively on the precise details of the distribution of errors in assembly.
\bibliographystyle{unsrt}

\begin{thebibliography}{10}

\bibitem{fratzl} Fratzl, P and Weinkamer, R. Nature's hierarchical materials.  \emph{Progress in Materials Science} \textbf{52}, 1263-1334 (2007).

\bibitem{meyers} Meyers, M. A., McKittrick, J. and Chen, P. Structural Biological Materials: Critical Mechanics-Materials Connections. \emph{Science} \textbf{339}, 773-780 (2013).

\bibitem{launey} Launey, M. E., Buehler, M. and Ritchie, R. O. On the Mechanistic Origins of Toughness in Bone. \emph{Annual Review of Materials Research} \textbf{40}, 25-53 (2010).

\bibitem{yao} Yao, H. and Gao, Huajian. Multi-scale cohesive laws in hierarchical materials. \emph{International Journal of Solids and Structures} \textbf{44}, 8177-8193 (2007).

\bibitem{lakes} Lakes, R. Materials with structural hierarchy. \emph{Nature} \textbf{361}, 511-515 (1993).

\bibitem{wegst} Wegst, U., Bai, H., Saiz, E., Tomsia, A. and Ritchie, R. Bioinspired structural materials. \emph{Nature Materials} \textbf{14}, 23-26 (2015).

\bibitem{jmax} Maxwell, J. C. On the calculation of the equilibrium and stiffness of frames. \emph{Philosophical Magazine} \textbf{27}, 294-299 (1864).

\bibitem{feng} Feng, S., Thorpe, M. F. and Garboczi, E., Effective-medium theory of percolation on central-froce elastic networks. \emph{Physical Review B} \textbf{31}, 276-280 (1985).

\bibitem{das} Das, M., MacKintosh, F. C., and Levine, A. J. Effective Medium Theory of Semiflexible Filamentous Networks. \emph{Physical Review Letters} \textbf{99}, 038101 (2007).

\bibitem{mao} Mao, X., Stenull, O. and Lubensky, T. C. Effective-medium theory of a filamentous triangular lattice. \emph{Physical Review E} \textbf{87}, 042601 (2013).

\bibitem{sen} Sen, D. and Buehler, M. J. Structural hierarchies define toughness and defect-tolerance despite simple and mechanically inferior brittle building blocks. \emph{Scientific Reports} \textbf{1}, Article 35 (2011).

\bibitem{hormoz} Hormoz, S., and Brenner, M. P. Design Principles for self-assembly with short-range interactions. \emph{Proceedings of the National Academy of Sciences} \textbf{108}, 5193-5198 (2011).

\bibitem{zeravcic} Zeravcic, Z., Manoharan, V. N., and Brenner, M. P. Size limits of self-assembled colloidal structures made using specific interactions. \emph{Proceedings of the National Academy of Sciences} \textbf{111}, 15918-15923 (2014).

\bibitem{dunlop} Dunlop, J. W. C., and Fratzl, P. Biological Composites. \emph{Annual Review of Materials Research} \textbf{40}, 1-24 (2010).

\bibitem{harris} Harris, J., B\"ohm, C. and Wolf, S. Universal structure motifs in biominerals:
a lesson from nature for the efficient design of bioinspired functional materials. \emph{Interface Focus} \textbf{7}, 20160120 (2017).

\bibitem{gab} Garboczi, E. J. and Thorpe M. F. Effective-medium theory of percolation on central force elastic networks. II. Further results. \emph{Physical Review B} \emph{31}, 7276-7281 (1985).

\bibitem{wilhelm} Wilhelm, J. and Frey, E. Elasticity of Stiff Polymer Networks. PRL 91, 108103 (2003).

\bibitem{fengj} Feng, J., Levine, H., Mao, X., and Sander, L. Nonlinear elasticity of disordered fiber networks. \emph{Soft Matter} \textbf{12}, 1419-1424 (2016).

\bibitem{head} Head, D. A., Levine, A. J., and MacKintosh, F. C. Distinct regimes of elastic response and deformation modes of cross-linked cytoskeletal and semiflexible polymer networks. \emph{Physical Review E} \textbf{68}, 061907 (2003).

\bibitem{lind} Lindstrom, S. B., Vader, D. A., Kulachenko, A., and Weitz, D. A. Biopolymer network geometries: Characterization, regeneration, and elastic properties. \emph{Physical Review E} \textbf{82}, 051905 (2010).

\bibitem{broed2} Broedersz, C. P. and MacKintosh, F. C., Modeling semiflexible polymer networks. Rev. Mod. Phys. 86, 995-1036 (2014).

\bibitem{amuasi} Amuasi, H. E., Heussingler, C. Vink, R.L.C. and Nonlinear and heterogeneous elasticity of multiply crosslinked biopolymer networks. \emph{New Journal of Physics} \textbf{17}, 083035 (2015).

\bibitem{licup} Licup, A. J., et al. Stress controls the mechanics of collagen networks. \emph{Proceedings of the National Academy of Sciences} \textbf{112}, 9573-9578 (2015).

\bibitem{lenz} Lenz, M. and Witten, T. A. Geometrical frustration yields fibre formation in self-assembly. \emph{Nature Physics} \textbf{13}, 1100-1105 (2017).

\bibitem{broed1} Broedersz, C. P., Mao, X., Lubensky, T. C., and MacKintosh, F. C. Criticality and isostaticity in fibre networks. \emph{Nature Physics} \textbf{7}, 983-985 (2011).

\bibitem{zhang} Zhang, T., Schwarz, J. M. and Das, M. Mechanics of anisotropic spring networks. \emph{Physical Review E} \textbf{90}, 062139 (2014)

\bibitem{lub} Lubensky, T.C.,Kane, C.L., Mao, X., Souslov, A. Sun, K. Phonons and elasticity in critically
coordinated lattices. \emph{Reports on Progress in Physics} \emph{78}, 073901 (2015).

\bibitem{fire} Bitzek, E., Koskinen, P., Gaehler, F., Moseler, M. and Gumbsch, P. Structural Relaxation Made Simple. \emph{Physical Review Letters} \textbf{97}, 170201 (2006).

\bibitem{JC} Phillips, J. C., Braun, R., Wang, W., Gumbart, J., Tajkhorshid, E., Villa, E., Chipot, C., Skeel, R. D., Kale, L., and Schulten, K. Scalable molecular dynamics with NAMD. \emph{Journal of Computational Chemistry} \textbf{26}, 16 (2005).




\end{thebibliography}

\end{document}